# Prediction of Coronary Heart Disease Using Routine Blood Tests

Ning Meng, Peng Zhang, Junfeng Li, Jun He and Jin Zhu[1]

**Background**—The objective of this study was to examine the association of routine blood test results with coronary heart disease (CHD) risk, to incorporate them into coronary prediction models and to compare the discrimination properties of this approach with other prediction functions.

**Methods and Results**—This work was designed as a retrospective, single-center study of a hospital-based cohort. The 5060 CHD patients (2365 men and 2695 women) were 1 to 97 years old at baseline with 8 years (2009–2017) of medical records, 5051 health check-ups and 5075 cases of other diseases. We developed a two-layer Gradient Boosting Decision Tree(GBDT) model based on routine blood data to predict the risk of coronary heart disease, which could identify 86% of people with coronary heart disease. We built a dataset with 15,000 routine blood tests results. Using this dataset, we trained the two-layer GBDT model to classify healthy status, coronary heart disease and other diseases. As a result of the classification after machine learning, we found that the sensitivity of detecting the health data was approximately 93% for all data, and the sensitivity of detecting CHD was 93% for disease data that included coronary heart disease. On this basis, we further visualized the correlation between routine blood results and related data items, and there was an obvious pattern in health and coronary heart disease in all data presentations, which can be used for clinical reference. Finally, we briefly analyzed the results above from the perspective of pathophysiology.

**Conclusions**—Routine blood data provides more information about CHD than what we already know through the correlation between test results and related data items. A simple coronary disease prediction model was developed using a GBDT algorithm, which will allow physicians to predict CHD risk in patients without overt CHD.

---

[1] Ning Meng, Peng Zhang and Junfeng Li are with Suzhou Institute of Advanced Study and School of Software Engineering, University of Science and Technology of China, Suzhou, China (E-mail: mengning@ustc.edu.cn, {sa517497, sa517167}@mail.ustc.edu.cn)

Jun He is with Suzhou Kowloon Hospital, Shanghai Jiaotong University School of Medicine, Suzhou, China (E-mail: hejun@kowloonhospital.com)

Jin Zhu is with Department of Urology, the Second Affiliated Hospital of Soochow University, Suzhou, China (E-mail: urologist.zhujin@gmail.com)

Corresponding author: Jin Zhu, urologist.zhujin@gmail.com; Jun He, hejun@kowloonhospital.com

## I. INTRODUCTION

Early and accurate identification of individuals with increased risk of coronary heart disease (CHD) is critical for effective implementation of preventative lifestyle modifications and medical interventions, such as statin treatment[1,2]. Thus, risk scores such as the Framingham Risk Score (FRS)[3], the American College of Cardiology/American Heart Association 2013 risk score (ACC/AHA13)[1] and genomic risk score (GRS) for CHD[4], which are based on clinical factors, lipid measurements and genome-wide association studies (GWAS), have been developed for CHD risk prediction. Although the scores can identify individuals at very high risk, a large proportion of individuals developing CHD during the next 10 years are not identified by these risk scores because they may be young, the tests are not affordable or they do not have access to sufficient healthcare to be evaluated by these risk scores.

It is very important to make full use of data in conventional medical service to identify high-risk populations with CHD using low-cost automation. This study shows that there is a significant correlation between CHD and blood components, and since routine blood routine examination is performed in most patients during routine medical service, it is simple and effective to evaluate the risk of CHD by using routine blood test data.

## II. MATERIALS AND METHODS

A total of 16860 patients were enrolled from eastern China; 5060 patients had CHD, including 5051 undergoing routine health check-ups and 5075 cases of other diseases.
The technical route consisted of clinical data collection, clinical data preprocessing, and feature engineering, and the data were merged into a health set, a diseased set, a CHD set and another disease set according to the diagnosis. Then, the machine learning algorithm model training, model evaluation, feature importance and correlation analysis were carried out.

**Clinical data collection:** According to the target population, from the relevant information system, we collected medical information of patients with CHD, including patients before admission from their outpatient examination data, the treatment period of collection of the examination data and other examination data. In addition, the health examination data of the healthy population and other patients' examination data were collected. These data included the patients' basic information, laboratory check data, diagnostic data, etc.

**Clinical data preprocessing**: Since the collected medical data usually have the characteristics of high volume, large sparsity and diverse data types, it is necessary to preprocess the data.

The preprocessing steps were as follows.

1) Data types were converted to the same type for different types of data, including the conversion of "male" in the gender data to 1 and the conversion of "female" to 0.

2) For data markers, the original data were divided into three categories: CHD patient data, other disease patient data and healthy population data, respectively labeled "1", "–1", and "0".

3) Data that are missing values are deleted to improve the effectiveness of data, because the collection of medical data sparsity is high. Then, the missing values are filled with the average value, specifically, the missing value of coronary heart disease is filled with the average of coronary heart disease, as are other diseases and health data.

4) Anomaly deletion was performed when the data in the process of production or transmission would produce an abnormal situation, combined with the business scene and specific rules to identify and remove the exception point.

5) Data integration was performed after the above steps; the remaining data totaled 15,033. To ensure the balance of training data, the data from 5,000 patients were randomly sampled from the CHD patient data, other disease patient data and healthy population data, and the data were consolidated by the 1:1:1 ratio to generate a complete dataset.

**Feature Engineering:** Statistics of the missing degree of each feature were performed before the data missing value was filled, because the healthy population underwent a health examination, causing the healthy population to have the highest data integrity. The main feature is the blood routine 22 examination index, which other data mostly lacks and therefore uses the basic information and the blood routine examination index to construct the feature set, which includes gender, age, WBC, RBC, MCHC, MONO, HCT, NEUT, MCV, PLT, HGB, LY, LY%, MONO%, NEUT%, RDW, BASO%, HgB, EOS, BASO, EOS%, MPV, PCT, and PDW.

**Selection and training of models:** according to the features of the dataset after preprocessing, we chose three different machine learning algorithms, LR (Logistic Regression), SVM (Support Vector Machines) and GBDT (Gradient Boosting Decision Tree). LR is a very common algorithm in the field of machine learning, which is widely used and is a classical classification model. SVM was proposed in 1995 by Corinna Cortes and Vapnik and other people looking for a hyperplane to classify the data, with relatively good performance indicators[9]. GBDT was presented in 2001 by Friedman[5,6] and is composed of Decision Tree Learning[7] and Gradient Boosting[8].

Our experiment is divided into two layers, the first layer is a health and disease two classification model, the second layer is for CHD and other diseases two classification model, using Mathieu Blondel's and other people's developed Sklearn library[14] for model training. The steps were as follows.

1) Dataset markup was performed in the first layer. The healthy population data were labeled "-1" and disease patient data labeled "1" (including patients with CHD and other diseases).
2) Dataset partitioning was performed in the first layer. Of the 15,000 marked data, 5,000 healthy people and randomly selected 5,000 patients were merged into a dataset, then we randomly sampled 70% of the data as a training set used for the training model, and the remaining 30% as a validation set for model effect evaluation.
3) Dataset markup was also performed in the second layer. Of the 15,000 marked data, 5,000 healthy population data from 15,000 data were eliminated, and the remaining CHD patient data were labeled "1", and the other patients were labeled "–1".
4) Dataset partitioning was then performed in the second layer. Of the labeled 10,000 data, 70% of the data were randomly sampled as a training set used for the training model, and the remaining 30% were used as a predictive set for model effect evaluation.
5) Model optimization was carried out for the specific data to adapt to different degrees, so the use of a grid search for parameter tuning was used for the prediction effect to achieve the optimum.
6) Model evaluation included evaluating the performance of the model, the diagnostic accuracy, sensitivity, specificity and precision were used to measure the diagnostic results. Using this model to diagnose the results of the predictive set, the results were compared with the known diagnostic results, and the accuracy, sensitivity and specificity of the model were obtained, and the specific meanings of these indexes are shown in Figure 1.

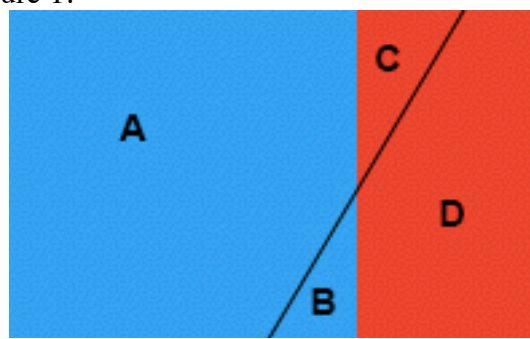

Figure 1. The meanings of the accuracy, sensitivity (recall), specificity and precision. The blue area means positive, the red area means negative, A is true positive, B is false negative, C is false positive, D is true negative. The accuracy is (A+D)/(A+B+C+D), sensitivity is A/(A+B), specificity is D/(C+D) and precision is A/(A+C)

TABLE-1: RESULTS OF THE HEALTH AND THE DISEASES, COMPARISON OF PERFORMANCE INDEXES OF THREE DIFFERENT ALGORITHMS FOR THE FIRST LAYER MODEL

| Algorithms | Accuracy | Sensitivity | Specificity |
|---|---|---|---|
| LR | 84% | 84% | 83% |
| SVM | 86% | 87% | 85% |
| GBDT | 93% | 93% | 91% |

TABLE-2: RESULTS OF CORONARY HEART DISEASE AND OTHER DISEASES, COMPARISON OF PERFORMANCE INDEXES OF THREE DIFFERENT ALGORITHMS FOR THE SECOND LAYER MODEL

| Algorithms | Accuracy | Sensitivity | Specificity |
|---|---|---|---|
| LR | 85% | 84% | 84% |
| SVM | 88% | 87% | 86% |
| GBDT | 91% | 93% | 90% |

The results of the experimental evaluation were obtained by training tuning in Table-1 and Table-2 for the three models. Three kinds of algorithms, whether of the accuracy, sensitivity or specificity, for GBDT were better. SVM is more sensitive to data and it is difficult to find a suitable kernel function, and GBDT does not need to consider this problem. In the fitting effect, GBDT is a combination of a weak classifier, not easy to occur with overfitting, and LR is easy to underfit. In the aspect of efficiency, SVM was the most time-consuming and had the lowest efficiency and GBDT the highest efficiency. To sum up, for all of the two layers we chose the GBDT algorithm. For the GBDT algorithm, if a logarithmic loss of function was similar to the logistic regression used, the loss of function is as follows.

$$L(y, f(x)) = \log(1 + \exp(-yf(x)))$$

where $y \in \{-1, +1\}$. Then, the negative gradient error is as follows.

$$r_{ti} = -\left[\frac{\partial L(y, f(x_i))}{\partial f(x_i)}\right]_{f(x) = f_{t-1}(x)} = y_i / (1 + \exp(y_i f(x_i)))$$

For the resulting decision tree, the optimal residual fitting value for each leaf node is as follows.

$$c_{tj} = \sum_{x_i \in R_{tj}} r_{ti} \Big/ \sum_{x_i \in R_{tj}} |r_{ti}|(2 - |r_{ti}|)$$

The optimization of parameters is achieved by using a grid search, in which GBDT is optimal when the learning rate is learning_rate=0.23 and the maximum number of iterations is n_estimators=70.

GBDT in the classification, regression and other machine learning tasks had very good results. It has also been proven to be a very efficient model in practice and is widely used.

VII. *CALCULATION*

The two-layer GBDT algorithms, which were trained together, could be combined into three classification models for health, CHD and other diseases. In fact, we could also use the GBDT three-classification model for direct training and evaluation, but in view of the reasons for the effective analysis of classification, the machine learning black box had to be opened to a certain extent to obtain blood routine and related data items of the association law to further understand the blood routine detection of the pathophysiology of data. At the same time, it is convenient to expand the classification evaluation to other diseases in the future, so we chose a two-layer GBDT model for the three classification, which is composed of two layers of GBDT algorithms, and the experimental results show that the two-layer GBDT model and GBDT three classification models are roughly the same in classification effect. In the above two-layer GBDT model, the evaluation results are as follows in Table-3.

TABLE-3: CALCULATION RESULTS OF TWO-LAYER GBDT MODELS

| Layers | Classes | Precision | Recall |
|---|---|---|---|
| First Layer | Healthy people | 87% | 91% |
| First Layer | Diseased people | 96% | 93% |
| Second Layer | Coronary heart disease (CHD) | 90% | 93% |
| Second Layer | Other diseases | 92% | 90% |

Calculation formula of CHD in three-classification model of two classification GBDT as follows.

Precision of CHD = Precision of first layer diseased people * Precision of second layer CHD

Recall of CHD = Recall of first layer diseased people * Recall of second layer CHD

According to these calculation formulas, the Precision and Recall of CHD are shown in Table-4.

TABLE-4: CALCULATION RESULTS OF CORONARY HEART DISEASE

| Classes | Precision | Recall |
|---|---|---|
| Coronary heart disease(CHD) | 86.4% | 86.5% |

## VIII. *RESULTS*

To carry out the risk assessment of coronary heart disease at a low cost, this paper established a two-layer GBDT classification model by means of machine learning and classification prediction of a large number of case data containing blood routine data. The results of the classification experiment showed that the model can identify approximately 86% of patients at high risk of CHD. This model provides a simple and low-cost approach to risk assessment of coronary heart disease.

**Feature importance and correlation analysis**

The GBDT algorithm is composed of multiple decision trees, and each tree's construction process is performing feature selection, through the specified metrics, selecting a feature among the candidate features and the corresponding split value; the tree level of the feature is closer to the root node, the more divided the number of times, the more important the features.

By using the Pearson method to analyze the correlation of all of the features, the two-layer GBDT classification model was sorted by the feature importance, among which three items with the highest contribution degree in the first layer model were RBC, HCT and LY%. The highest contribution to the second layer model in addition to age were PDW, RDW and MPV. We selected several features of higher contribution to the healthy population and coronary heart disease population from two contributions ranking, the selected features including LY%, HCT, RBC, age, RDW, BASO% and LY for correlation analysis, as shown in Figure 2. The population of healthy people and coronary heart disease crowd dot matrix aggregation effect is obvious, and among them, the blue is the healthy crowd, red is the coronary heart disease crowd, and obviously the healthy crowd's aggregation effect is better than the coronary heart disease crowd. This along with our result for the healthy population recall rate of 91% is higher than the coronary heart disease recall rate of 86.5%.

Furthermore, we present a visual representation of all features, blood routines and their associated data items, to further identify the complex relationships between these features and the regularity of aggregation. We connected all of the features with a dashed line, as shown in Figure 3, green for healthy people, red for coronary heart disease and blue for other diseases, and the green and red aggregation is clearly visible, which confirms the classification results of our model, and visualization presents its intrinsic reasons.

## IX. *DISCUSSION*

At present, the incidence of cardio-cerebrovascular disease is increasing rapidly, and the high mortality, morbidity and relapse rates of coronary heart disease lead to a heavy social and economic burden. At present, China's current cardiovascular disease patients number at least 290 million, including myocardial infarction patients approximately 2.5 million, and each year, approximately 3.5 million people die of cardiovascular disease, the highest death toll among all diseases.

Coronary heart disease has a long period of onset, a slow progress, and no obvious symptoms until it becomes quite advanced. Once the onset of physical discomfort and other symptoms manifest, the disease has been present for many years, there is a life-threatening risk, and the cost of treatment is high.

The current diagnosis of CHD is mainly based on coronary angiography, but it is obvious that the disease has to become quite serious before it is evident on coronary angiography. In view of the high incidence of coronary heart disease, its threat to health, and the treatments used for the disease, for early screening for CHD there are currently no good methods. Therefore, it is necessary to identify a simple and effective early screening method.

Our findings provide a scientifically effective, low-cost screening tool. The blood routine index can predict the risk of coronary heart disease so well that we believe there must be a deep pathophysiological basis behind it. In fact, several characteristics of coronary heart disease changes in hemorheology, inflammatory response, and coagulation function, which can be reflected in the blood routine indicators. We selected the first three indexes (coronary heart disease, healthy person and CHD, other disease Prediction Index) to review the literature and found that these indexes were reliable indexes of coronary heart disease prediction and had their own associated pathophysiology.

**RBC and coronary heart disease.** The results showed that RBC was significantly higher in the coronary heart disease group than in the control group. Coronary heart disease causes the organism to be in anoxic condition, and long-term hypoxia stimulates RBC and HGB compensatory generation to meet the increased needs of hypoxia; however, too many RBC and HGB promote the patient's high viscosity of blood platelets in the arterial wall adhesions, red blood cell agglutination ability is enhanced and erythrocyte deformability is decreased, and thus the whole blood viscosity is increased. The blood flow resistance increases in the coronary artery and stenosis can easily result in thrombosis, causing myocardial infarction.

**HCT and coronary heart disease.** This study found that the HCT of coronary heart disease was higher than in the normal group although the specific mechanism is not very clear. It is generally believed that the whole blood viscosity is an important factor to determine blood flow, and HCT is one of the determinants of whole blood viscosity. HCT increases in red blood cell aggregation, resulting in increased viscosity. Blood viscosity is often caused by the microcirculation of blood flow being slowed down and oxygen transport decreased, resulting in circulatory hypoxia, local hypoxia, and a high coagulation state. In addition, high levels of HCT are often associated with other risk factors such as BMI, hypertension, high cholesterol, impaired glucose tolerance, smoking and so on, so the risk of coronary heart disease is higher in HCT patients.

**Lymphocytes and coronary heart disease.** The results showed that the ratio of lymphocytes in patients with coronary heart disease decreased and the ratio of neutrophil/lymphocyte was higher than normal. However, the mechanism is not entirely clear. Some people think that a reduction in lymphocyte count may be due to physiological stress caused by cortical hormone secretion and an immune response that is out of control.

**PDW and coronary heart disease.** The results showed that the PDW value of CHD patients was significantly higher than that in the healthy control group. Platelets are an important factor in initiating thrombosis, and a large volume of platelet function plays an important role in the development of coronary heart disease. PDW is a parameter that reflects the heterogeneity of platelet size. The important pathological changes in coronary heart disease are thrombosis, and the coagulation activity enhanced therein plays an important role, manifesting in platelet adhesion and gathering, and the release reaction enhances and causes thrombosis because of the massive platelet consumption. The marrow giant nucleus cell feedback multiplication is because the volume is too large, so the young platelets enter the peripheral blood, and the massive volume of platelet activity is strong. Released medium further participates in the dynamic evolution of the atherosclerotic plaque because the average platelet volume increases so that the size of platelets is increased, and the PDW is also increased.

**RDW and coronary heart disease.** A large number of studies have confirmed that the RDW index of CHD patients is higher than among those in normal health. This index partly reflects the severity of myocardial damage, and the higher the RDW level of patients, the worse the outcome. RDW and some inflammatory markers such as n-terminal natriuretic peptide precursor and C-reactive protein have a certain correlation. It is believed that RDW inflammation activity is increased and neuroendocrine system activation is closely related. Another mechanism may be that, because of inflammatory cytokines, bone marrow red-line stem cells are insensitive to erythropoietin stimulation, preventing their resistance to apoptosis and promoting maturation. Inflammatory cytokines inhibit the maturation of red blood cells, resulting in immature red blood cells entering the circulation, increasing the size and heterogeneity of red blood cells, and raising the RDW.

**MPV and coronary heart disease.** MPV is the index used for evaluating platelet function and activity, and the size of MPV reflects the proliferation and platelet formation of giant nucleus cells in the bone marrow and is closely related to platelet life, ultrastructure and their functional status in circulating blood, and a decrease in platelet volume indicates that their physical substance, activity and function are decreased. An increase in MPV had a significant positive correlation with a reduced bleeding time, an increase in platelet A2 release, and the expression of the platelet membrane glycoprotein Ib and the IIb/IIIa receptor. Coronary heart disease patients have blood in a high coagulation state, and thus the platelets find it easier to adhere to the accumulation and induce activation, forming a cycle of platelet relative reduction, and thus promoting a reaction of the bone marrow giant cells to produce large volumes of platelets, so that the MPV increases.

We note that there are significant differences in the ranking of indicators for the two layers of health/disease and coronary heart disease/other diseases. We speculate on possible causes: Many other diseases also have a certain effect on routine blood tests, so that some specific indicators may change in the direction of coronary heart disease or the contrary, so that the contribution of a certain indicator may be more significant or not significant.

We also note that the scatter plot shown in Figure 2 in healthy people's indicators are relatively concentrated, while the coronary heart disease patients are relatively dispersed indicators; there is a famous saying:

*All happy families are all alike; every unhappy family is unhappy in its own way.*
   *--Leo Tolstoy, Anna Karenina*

Obviously, the healthy people's index similarity is relatively high, and the patient's condition is different. The index is relatively dispersed, but its overall or relative concentration has certain laws to follow. This is the reason why our model predicts better results.

## X. CONCLUSIONS

We developed a two-layer GBDT model based on routine blood test data to predict the risk of coronary heart disease, which can identify 86% of people with coronary heart disease. We built a dataset with 15,000 routine blood test data. On this dataset, we trained several machine learning models to classify health, coronary heart disease and other diseases. As a result of the classification of machine learning, we found that the sensitivity of detecting the health data was approximately 93% in all data, and the sensitivity of detecting CHD was 93% in the diseases data, that included coronary heart disease. On this basis, we further visualized the correlation between routine blood tests and related data items, and there was an obvious regularity for health and coronary heart disease in all of the data presentations, which can be used for clinical reference. Finally, we briefly analyzed the pathophysiology of these results from the perspective of pathophysiology.

*ACKNOWLEDGMENTS*

This article required many people's help, support and direct or indirect contributions from the idea to the written article, including Li Hua, Fu Hong, Xiong Yaping, Xiang Jie, Kou Yafei, Yuan Guo, Tian Qi, Wang He, Lin Jiafeng and all of the students in the np2016 class and the DoctorTalent Lab.

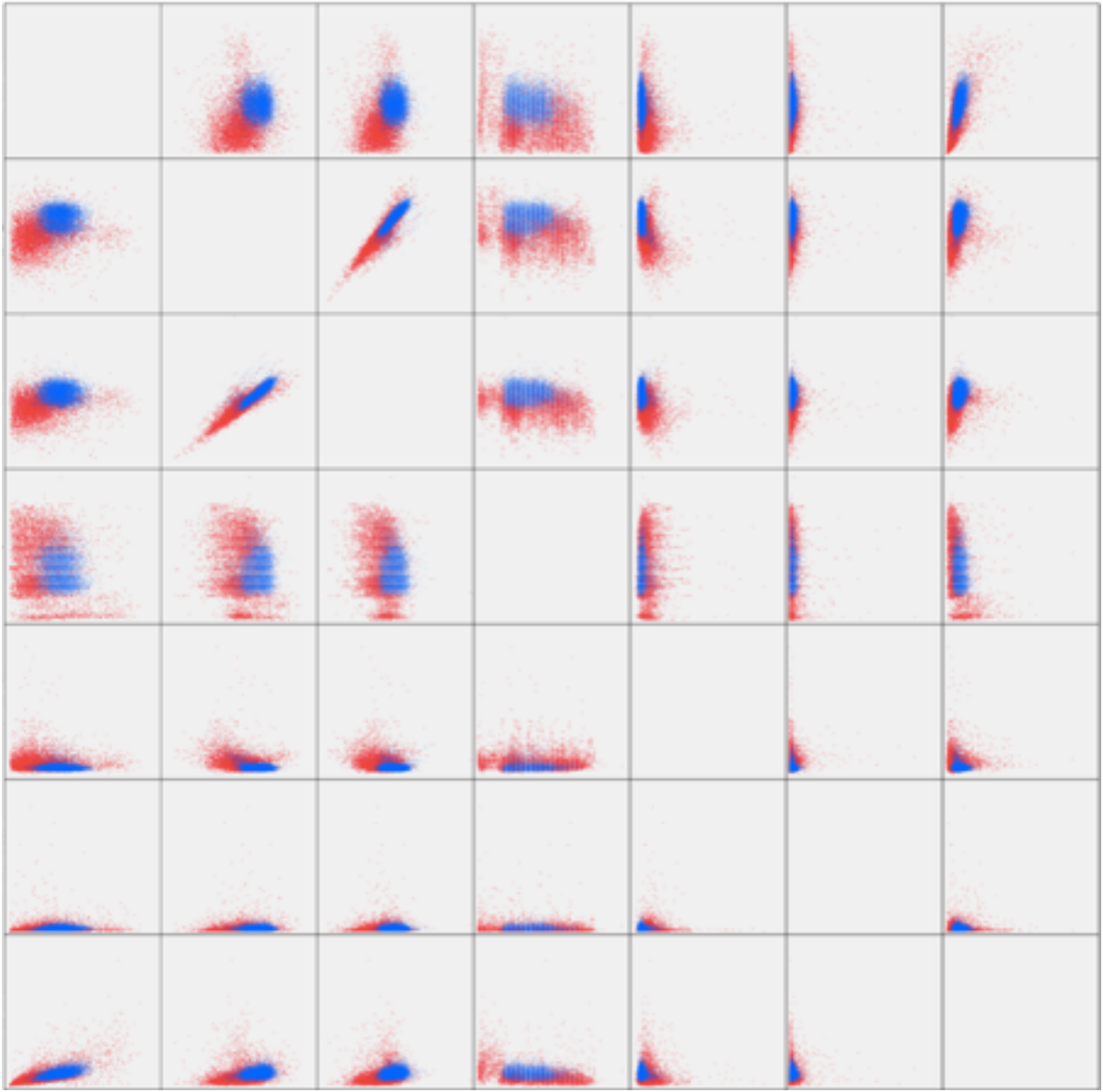

Figure 2. Correlation between LY%, HCT, RBC, age, RDW, BASO% and LY. The population of healthy people and coronary heart disease crowd dot matrix aggregation effect is obvious. Among them the blue is the healthy crowd, red is the coronary heart disease crowd, and obviously the healthy crowd's aggregation effect is better than the coronary heart disease crowd.

15.

16.

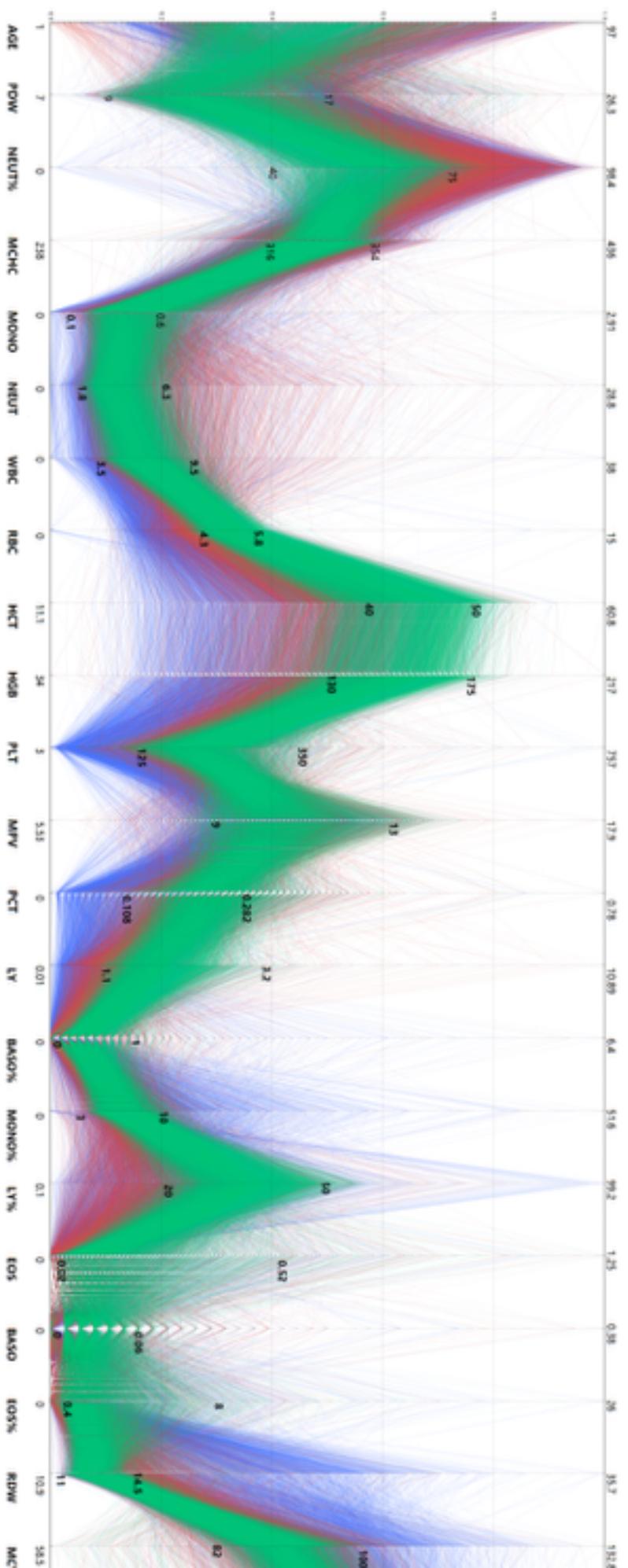

Figure 3. Visual analysis of features correlation: green for healthy people, red for coronary heart disease and blue for other diseases. Green and red aggregation is clearly visible, which confirms the classification results of our model, the visualization presents its intrinsic reasons.